\documentclass[twocolumn,superscriptaddress,aps,preprintnumbers,amsmath,amssymb,prl,nofootinbib]{revtex4}

\usepackage{graphicx}
\usepackage{epstopdf}
\usepackage{dcolumn}% Align table columns on decimal point
\usepackage{bm}% bold math
\usepackage{hyperref}
\usepackage{color}

\def\m{\mu}

\def\s{\sigma}

\def\G{\Gamma}

\def\beq{\begin{eqnarray}}
\def\eeq{\end{eqnarray}}

\newcommand{\vev}[1]{ \left\langle {#1} \right\rangle }

\begin{document}
\

\title{
Hypercharged Dark Matter and  Direct Detection as a Probe of Reheating
}

\author{Brian Feldstein}
\affiliation{Rudolf Peierls Centre for Theoretical Physics, University of Oxford, 1 Keble Road, Oxford, OX1 3NP, UK}
\affiliation{IPMU, University of Tokyo, Kashiwa, 277-8568, Japan}
\author{Masahiro Ibe}
\affiliation{IPMU, University of Tokyo, Kashiwa, 277-8568, Japan}
\affiliation{ICRR, University of Tokyo, Kashiwa, 277-8582, Japan}
\author{Tsutomu T.~Yanagida}
\affiliation{IPMU, University of Tokyo, Kashiwa, 277-8568, Japan}

\begin{abstract}
The lack of new physics at the LHC so far weakens the argument for TeV scale thermal dark matter.  On the other hand, heavier, non-thermal dark matter is generally difficult to test experimentally.  Here we consider the interesting and generic case of hypercharged dark matter, which can allow for heavy dark matter masses without spoiling testability.  Planned direct detection experiments will be able to see a signal for masses up to an incredible 
$10^{10}$\,GeV, and this can further serve to probe the reheating temperature up to about $10^9$\,GeV, as determined by the non-thermal dark matter relic abundance.
The $Z$-mediated nature of the dark matter scattering may be determined in principle by comparing scattering rates on different detector nuclei, which in turn can reveal the dark matter mass.  We will discuss the extent to which future experiments may be able to make such a determination.

\end{abstract}

\date{\today}
\maketitle
\preprint{IPMU13-0206}
\preprint{ICRR-report-2013-13}
\preprint{OUTP-13-22p}

The success of the Minimal Standard Model to date weakens the case for TeV scale thermal relic dark matter.  For heavier dark matter, however, testability becomes a key issue.  Interestingly, if dark matter is in some representation of $SU(2)_L$, along with a non-zero hypercharge to make one of its components electrically neutral, 
then tree level $Z$ exchange leads to significant cross sections at direct detection experiments, and can lead to signals for even very heavy masses.

In this letter, we discuss the possibility of hypercharged minimal dark matter \cite{MD}
produced non-thermally from the thermal-bath during the reheating process after inflation.
We will show that a signal at direct detection experiments would be correlated to concrete information about the reheating temperature and associated thermal history.  In particular, a signal at future experiments could effectively measure the reheating temperature to within a two order of magnitude window.  Planned detectors will be sensitive to masses of up to about $10^{10}$\,GeV, and in turn reheating temperatures of up to about $10^7$--$10^9$\,GeV.   Making such constraints compelling would require gaining evidence that heavy hypercharged dark matter was indeed responsible for an observed signal.  We will show that such evidence could be gleaned by comparing rates and spectra at multiple nuclear targets, to determine the $Z$-mediated nature of the scattering.  In particular, currently planned experiments have the capability to rule out a future signal as being mediated by Higgs exchange, or other isospin conserving possibilities, at almost 90\%\,C.L.  Hidden photon mediated scattering could be even more tightly constrained.  
Indeed, in such a situation, $Z$-mediated scattering might be the most compelling possibility, with farther off experiments capable of giving further evidence.

Interestingly, the scenario we discuss might arise in supersymmetric theories if the higgsino is the lightest superpartner, and if the supersymmetry breaking scale is very high-- as presently motivated by the lack of evidence for superpartners near the weak scale.

%There, we show that through the interrelation between the dark matter mass 
%and the reheating temperature after inflation required for the correct abundance,
%it is possible to probe the reheating temperature via direct detection experiments of dark matter.
%In particular, a  robust upper limit on the reheating up to ${\cal O}(10^9)$\,GeV
%will be placed once the dark matter signals are observed in the future experiments, 
%which excludes or lends support to  thermal leptogenesis\,\cite{leptogenesis}.
%We also discuss how to test the hypercharged minimal dark matter scenario
%by observing the isospin violation at the direct detection experiments.
%We also discuss that the $Z$-mediated nature of the direct detection interactions could 
%be discerned by comparing the nuclear recoil energy spectra on the different target materials.

\vspace{4pt}
%%%%%%%%%%%%%%%%%%%%
{\bfseries\slshape  Direct detection.}\/
In order for the dark matter $SU(2)_L$ multiplet to contain an electrically neutral particle  (which is then automatically the lightest \cite{MD}), a variety of hypercharge assignments are possible.  
For a doublet, for example, the hypercharge must be $1/2$.
If the hypercharge is non-zero,  the dark matter particle interacts with nuclei via $Z$-boson exchange, 
with a spin independent scattering cross section,
\begin{equation}
\label{eq:DD}
\sigma_{\chi N} = \frac{G_F^2 \mu_N^2}{2\pi} Y^2 (N - (1-4\sin^2\theta_W)Z)^2\ .
\end{equation}
where $G_F$ is the Fermi constant, $Y$ the dark matter hypercharge, $\m_N$ the reduced mass of the nucleus and dark matter, $\theta_W$ the weak mixing angle,
and $N$ and $Z$  the number of neutrons and protons in the target, respectively.  
Here, we have assumed fermionic dark matter. For the scalar case, the cross section is multiplied by a factor of 4.

The strongest direct detection constraint presently comes from the XENON\,100 experiment\,\cite{Aprile:2012nq}, 
and at large masses, it takes the form
\begin{eqnarray}
%\s_n \lesssim2\times 10^{-44}{\rm cm}^2 \times
\sigma_{\chi \rm{Xe}}\lesssim 6\times 10^{-36} {\rm cm}^2
\left( \frac{M_{\rm DM}}{1\,{\rm TeV}}\right),
\end{eqnarray}
at  90\%\,C.L, 
%assuming a local dark matter density, $v_0=0.3$\,GeV/cm$^3$,
%a local circular velocity of $v_0 = 220$\,km/s, and Galactic escape velocity of $v_{\rm esc} = 544$\,km/s.
%Here, $\sigma_n$ is  the dark matter nucleon cross section, taken as equal for protons and neutrons.
%This translates into the requirement $\sigma_{\chi \rm{Xe}}\gtrsim 6\times 10^{-36} {\rm cm}^2$ 
%for $m_{DM}=1$\,TeV,
 from which follows 
\begin{equation}
M_{DM} \gtrsim (2 Y )^2 \times 3 \times 10^7\,{\rm GeV}\ .
\label{eq:bound}
\end{equation}
It is remarkable that direct detection experiments are continuing to explicitly search for physics at such high mass scales.%
\footnote{It is possible to avoid the constraint of Eq.\,(\ref{eq:bound}) if the masses of the neutral components of the dark matter multiplet are split by more than the energy available in nuclear scatterings, ${\cal O}(100)$\,keV.  
This can be accomplished by mixing the dark matter multiplet with other particles, 
as is usually the case for the Higgsinos in the supersymmetric standard model.
However, this cannot be accomplished without adding non-minimal structure to the theory. 
%In the minimal case, satisfying direct detection constraints requires the dark matter particle to not be a standard thermal relic.
}

The constraint of Eq.\,(\ref{eq:bound}) is often taken to imply that 
the hypercharged minimal dark matter is strongly disfavored.  
In particular, the annihilation cross section of hypercharged minimal dark matter
into Standard Model particles is given by\,\cite{MD} (again, assuming a fermion)
\begin{eqnarray}
\label{eq:ann}
 \vev{\sigma v} &\simeq& \frac{1}{512\pi k M_{\rm DM}^2} 
\times (g_2^4(2k^4+17k^2 -19) \nonumber\\
&& + 4 Y^2 g_Y^4(41 + 8 Y^2) +16 g_2^2 g_Y^2 Y^2 (k^2 -1))\ , %\\
%&\simeq& 10^{-26}{\rm cm}^3/s \times \left(\frac{2\,{\rm TeV}}{m_{\rm DM}}\right)^{2}\ ,
\label{freeze}
\end{eqnarray}
where k is the dimension of the dark matter $SU(2)_L$ representation. 
Therefore, the standard thermal relic cross section,  ${\cal O}(10^{-26})$\,cm$^3$/s, is achieved for  $M_{\rm DM} = {\cal O}(1)$\,TeV, which clearly  contradicts with the above constraint.\footnote{The annihilation cross section is enhanced in the
presence of charged $SU(2)_L$ partners with almost degenerate 
masses\,\cite{Hisano:2002fk,Cirelli:2007xd},
although dark matter of $O(10)$\,PeV is still too heavy to lead to a correct thermal relic density
even with the enhanced annihilation cross section.%%%%%
}

\vspace{4pt}
%%%%%%%%%%%%%%%%%%%%
{\bfseries\slshape  Non-thermal production.}\/
On the other hand, even if one makes no additional assumptions beyond a standard hot early universe preceded by inflation, %then, as one adjusts only the dark matter mass, 
there are actually two values for the dark matter mass which lead to the correct relic density, not one.  
The first is of course the above mentioned thermal freeze-out mass, determined by Eq.\,(\ref{freeze}).
The second corresponds to taking a heavy dark matter mass, larger than the reheating  temperature 
of the universe after inflation, so that dark matter never attained equilibrium. 
This is the so-called WIMPZILLA scenario\,\cite{WIMPZILLA}.
There, the correct relic density is realized by carefully arranging the dark matter mass, the  maximum temperature
of the universe after inflation, $T_{\rm max}$, and  the reheating temperature $T_R$.
The first of these possibilities is generally considered more attractive primarily because it has been 
expected that new physics will be present at the TeV scale in any case to stabilize the Higgs mass.  
The second scenario, on the other hand, appears somewhat fine tuned due to the carefully chosen 
Boltzmann factor.
However, the present experimental situation suggests a rethinking of these arguments 
(see also discussions in the final section).
In the following, we consider the second scenario and discuss to what extent we can probe 
the reheating process via direct detection experiments.

The Boltzmann equation of the number density $n$ of dark matter is given by,
\begin{eqnarray}
\frac{d}{dt} n + 3 H n=-  \vev{\s v}(n^2 - n_{\rm EQ}^2) \ ,
\end{eqnarray}
where $n_{\rm EQ}$ denotes the thermal equilibrium number density.%
\footnote{We assume that dark matter is produced primarily through scattering in the thermal bath.  This essentially requires that the inflaton mass is lighter than the dark matter mass, so that direct inflaton decays to the dark matter particle are forbidden.}
%\begin{eqnarray}
%n_{\rm EQ} = 2 \left( \frac{mT}{2\pi}\right)^{3/2} e^{-\frac{m}{T}}\ .
%\end{eqnarray}
After the end of inflation
the Hubble parameter and the temperature of the universe 
depend on the scale factor $a$ as%
\footnote{
%Here, we are assuming a rather slow decay of the inflaton so that
%the back-reaction from the thermal bath is irrelevant.
When back-reaction is significant,
the effective exponent in the temperature scaling can be smaller\,\cite{Kolb:2003ke,Yokoyama:2005dv,Mukaida:2012bz}.
The following analysis can be extended to such cases straightforwardly, and our results would be qualitatively unchanged.
}
\begin{eqnarray}
H = H_R \left(a/a_R \right)^{- 3/2}\ , \quad
T = T_R \left(a/a_R \right)^{- 3/8}.
\end{eqnarray}
%where $\eta = 2$ and $\e = 1$ in the radiation dominated era
%while $\eta = 3/2$ and $\e = 3/8$ in the inflaton dominated era%
Here, the subscripts $R$ denote the values at the end of the reheating process.
The maximum temperature $T_{\rm max}$ may be related to the Hubble parameter at the end of inflation via
\begin{eqnarray}
 H_{\rm inf} \simeq H_R \times \left(T_{\rm max}/T_R \right)^4\ .
\end{eqnarray}

\begin{figure}[tb]
\begin{center}
  \includegraphics[width=0.75\linewidth]{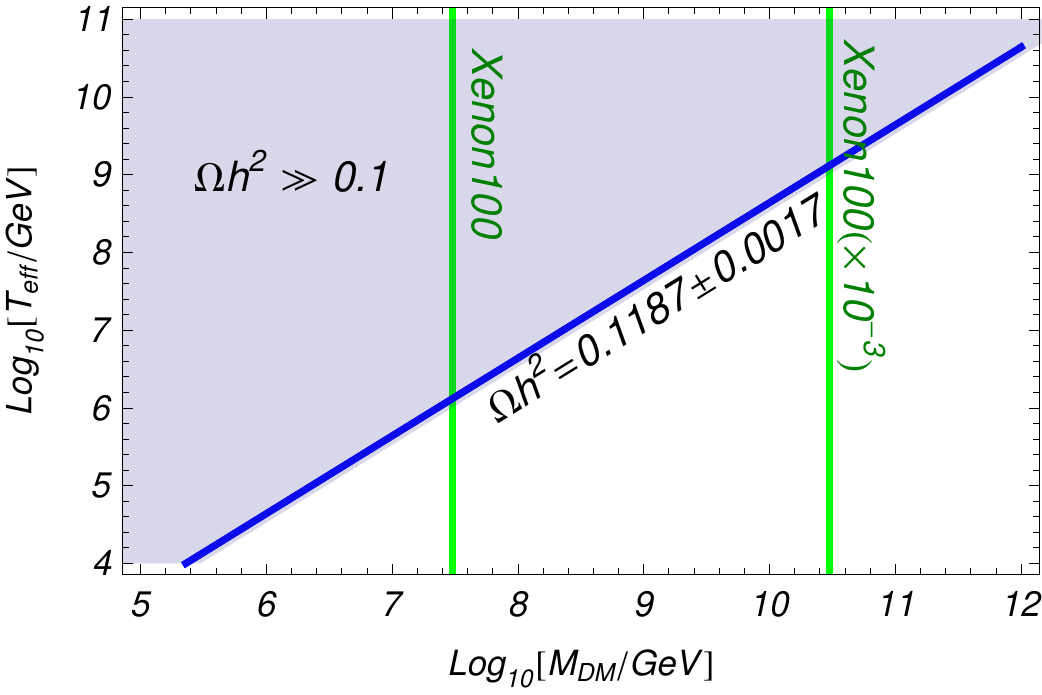}
\end{center}
\caption{\sl \small
Contour plot of the relic density of minimal dark matter for a $Y=\frac{1}{2}$  doublet fermion.
The thick (blue) line corresponds to the observed dark matter density.
The correct density is realized for $M_{\rm DM}/T_{\rm eff} \simeq 23$ independent 
of $M_{\rm DM}$.
The vertical (green) lines show the current lower limit on the mass from the XENON\,100 experiment and the projected
constraint after a thousand times increased sensitivity.
}
\label{fig:Teff}
\end{figure}

The dark matter abundance is straightforwardly obtained by solving the above equation 
with initial condition $n_X(T_{\rm max}) = 0$.
As a result, we obtain the resultant relic density, for e.g. a $Y=1/2$ doublet,
\begin{eqnarray}
\Omega_{\rm DM}h^2 \simeq 
 \frac{1}{36 \pi^6} \left( \frac{45}{2g_*}\right)^{3/2} \frac{s_0\vev{\s v} M^2}{H_0^2M_{\rm pl} }  e^{-2 x_{\rm eff}}\ ,
\end{eqnarray}
where $g_*\simeq {\cal O}(100)$ denotes the effective number of relativistic degrees of freedom during
the reheating process, $M_{\rm pl}$ the reduced Planck scale, $s_0$ the entropy density of the present 
universe, and $H_0 = 100$\,km/s/Mpc$^{-1}$.
The exponent, $x_{\rm eff}$, 
which stems from the Boltzmann suppression factor in $n_{EQ}$ is given by,
\begin{eqnarray}
\label{eq:xeff1}
x_{\rm eff} &=&
- 0.5\log\left[\Gamma(9, 2 x_{\rm max})\right] + 3.5 \log x_R. \label{xeff}
%\nonumber\\
%&&-\frac{1}{2}\log\left[k^4 \e^{-1}2^{-2 x_{\rm med}'} \G[2x_{\rm med}',2x_{\rm max}]\right]\ ,
%\end{eqnarray}
%for  $ \max[x_{\rm med},x_{\rm max}] < x_{R}$ and
%\begin{eqnarray}
%x_{\rm eff} = x_R- \frac{1}{2} \log\left[ k^4 x_R \right]\ ,
\end{eqnarray}
%for  $x_{\rm med} > x_{R}$.
Here, we have introduced the variable $x = M_{\rm DM}/T$, and we note
%  and defined
%\begin{eqnarray}
%x_{\rm med} = \frac{3+\eta-4\e}{2\e}\ , \quad x_{\rm med}' = \frac{3+\eta-3 \epsilon}{2\epsilon}\ ,
%\end{eqnarray}
%$x_{\rm med} = (3+\eta-4\e)/{2\e}$ and $x_{\rm med}' = (3+\eta-3\e)/{2\e}$. 
%In the inflaton dominated period, we have $x_{\rm med} = 4$ and $x_{\rm med}' = 4.5$. 
that the relic density depends on the dark matter mass only through $x_{\rm eff}$.
Thus, we find that the observed dark matter density for
 $Y= \frac{1}{2}$ and
 $k =2$ is obtained with $x_{\rm eff} \simeq 23$ independent
of the dark matter mass.
%and with only a minor logarithmic dependence on
%$Y$ and
%$k$ for other representations.
(See Fig.\,\ref{fig:Teff}).

\begin{figure}[tb]
\begin{center}
  \includegraphics[width=0.75\linewidth]{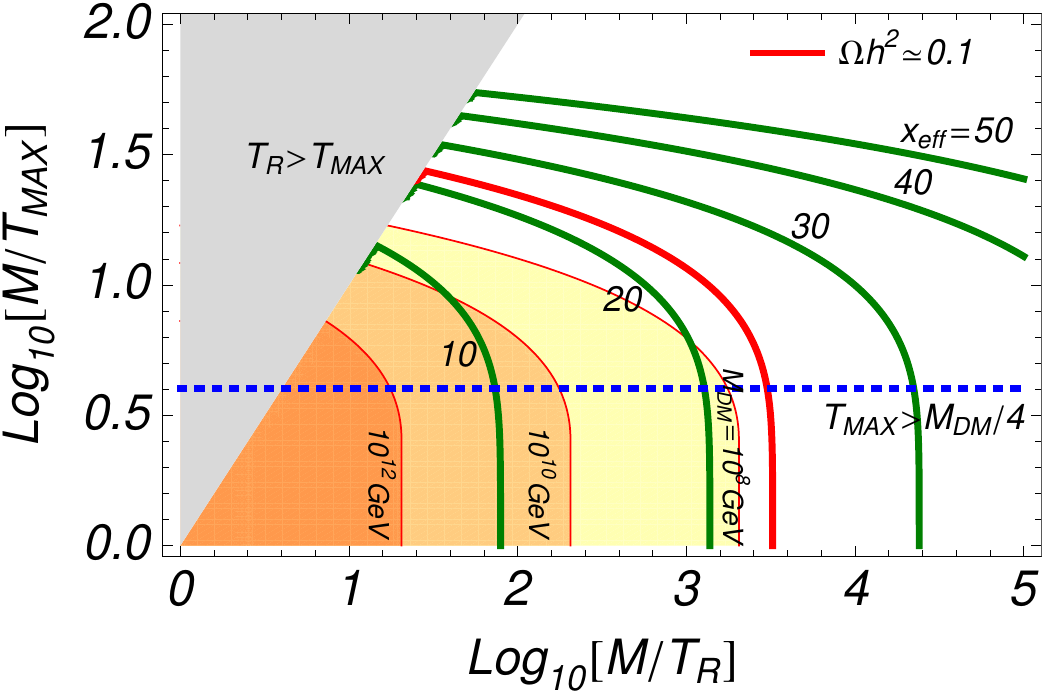}
\end{center}
\caption{\sl \small
Contour plot of $x_{\rm eff}$ in the $(T_R,T_{\rm max})$ plane.
The red line corresponds to $x_{\rm eff} \simeq 23$ for which the relic density 
reproduces the observed density.
In the orange shaded regions, dark matter attained thermal equilibrium in between
$T_{\rm max}$ and $T_R$ for a given dark matter mass.}
\label{fig:TRTmax}
\end{figure}

Now, let us discuss in detail the implications for $T_{\rm max}$ and $T_{R}$ of a positive observation of dark matter at upcoming direct detection experiments.
%When the maximal reheating temperature is much lower than $T_{\rm med}$, i.e. 
%for $x_{\rm med} \ll x_{\rm max}\ll x_{R}$,
%the effective temperature is reduced to,
%\begin{eqnarray}
%x_{\rm eff} \simeq   x_{\rm max}  - x_{\rm med} \log( {x_{\rm max}}/{x_R} )- \frac{1}{2} \log x_R/\e  \ .
%\end{eqnarray}
%In this case, the relic abundance depends exponentially 
%on $T_{\rm max}$, i.e. $\Omega_{\rm DM}h^2\propto e^{-2x_{\rm max}}$.
%For the very high maximum temperature, i.e. $x_{\rm max}\ll 1$, on the other hand,
%the effective temperature $T_{\rm eff}$ becomes independent of $T_{\rm max}$,
%and the relic density has no exponential dependences on $T_{\rm max}$ nor $T_{R}$\,\cite{WIMPZILA}.
In Fig.\,\ref{fig:TRTmax}, we show the parameter region which reproduces the observed 
dark matter density in the $(M_{\rm DM}/T_R, M_{\rm DM}/T_{\rm max})$ plane, i.e. we show the contour $x_{\rm eff}\simeq 23$.
The figure shows the interesting result that the observed dark matter density requires the reheating temperature to fall within a specific two order of magnitude window,
\begin{eqnarray}
30 \lesssim M_{\rm DM}/T_R \lesssim  10^{3.5}\ .
\end{eqnarray}
We note that the relic density becomes insensitive 
to $T_{\rm max}$ for $T_{\rm max} \gtrsim M_{\rm DM}/4$ for which $\G[9, 2x_{\rm max}]$
reduces to $\G[9]$ in Eq.\,(\ref{eq:xeff1}).
For  $T_{\rm max} \ll M_{\rm DM}$, on the other hand, the exponent $x_{\rm eff}$
has a linear dependence on $x_{\rm max}$, and hence, the relic density 
depends on the maximal temperature exponentially.  We thus see that there are two qualitative
regimes in which the correct relic abundance may be obtained; the first is by having a small maximum temperature, such that the final abundance is appropriately Boltzmann suppressed; the second is by having a larger maximum temperature, and then demanding that $T_R$ is much smaller than $M_{\rm DM}$ in order to obtain entropy production before reheating.

In Fig.\,\ref{fig:TRTmax}, we have also shown the parameter regions in which the
dark matter particle attained thermal equilibrium, i.e. $n_{\rm EQ}\vev{\s v}/H \gtrsim 1$, 
in between $T_{\rm max}$ and $T_R$.
We note that for essentially all masses of interest, hypercharged doublet dark matter never attained equilibrium anywhere along the $\Omega h^2 =.1$  contour 
even for very high maximum temperatures.  For other representations, the required value of $x_{\rm eff}$ obtains a mild logarithmic sensitivity to
$Y$ and $k$.  Moreover, for small $x_{\rm max}$ and larger representations, it is possible for DM to have obtained thermal equilibrium before reheating, which can somewhat change the form of the resulting relic abundance and therefore the expression (\ref{xeff}) for $x_{\rm eff}$.  In such cases however, we find that the lower bound on $T_R$ is slightly raised, with the upper bound being unaffected, so that the allowed window actually becomes smaller.

As we have emphasized above, direct detection experiments have placed a very stringent limit
on the dark matter mass in this scenario, $M_{\rm DM} \gtrsim (2 Y)^2 \times  3\times 10^7$\,GeV, and planned
experiments will probe up to masses of $10^{10-11}$\,GeV\,\cite{Baudis:2012ig}.
Therefore, if a dark matter signal were observed, 
the allowed reheating temperature in this model would then be constrained to
\begin{eqnarray}
T_R \simeq10^{7-9}\,{\rm GeV}  \left(\frac{M_{\rm DM}}{3\times10^{10}\,{\rm GeV}}\right)\ .
\end{eqnarray}
Thus, in the hypercharged dark matter scenario, it is possible
to probe very high temperatures of the universe through direct detection experiments.  This would have interesting implications in particular for thermal leptogenesis \cite{leptogenesis}, which could be excluded by a low reheating temperature, or perhaps lent some support by a high reheating temperature.

Let us briefly note that  super heavy dark matter can also be gravitationally produced 
in the transition between the inflationary to the inflaton dominated epoch\,\cite{Chung:1998zb}.
We find, however, that the non-thermal production from the thermal bath
dominates for $M_{\rm DM} \lesssim 10^{11}$\,GeV for the scenario we consider here.
Therefore, the above discussion is not altered within the mass range 
which is relevant for planned direct detection experiments.

%%%%%%%%%%%%%%%%%
\vspace{4pt}
{\bfseries\slshape  Testability.}\/
As given in Eq.\,(\ref{eq:DD}), the spin independent nuclear scattering cross section in this scenario
 shows the isospin violating nature of the $Z$-boson exchange interactions.

Here, we discuss to what extent we could explicitly test this scenario by measuring $f_p/f_n$, the ratio of the proton
and neutron couplings, which here takes the value $-(1 - 4 \sin^2 \theta_W) \sim -0.04$.  For that purpose, it would be crucial to compare the signal strengths obtained at multiple experiments having target materials with different ratios of protons and neutrons.  To distinguish different  values of $f_p/f_n$, a certain amount of statistical power would be necessary, and thus we consider the largest of the presently planned experiments, which will involve Xe, Ge, and Ar.  Following \cite{Pato:2011de}, we will take future Xe experiments to
have no background events within a recoil energy range of $10$--$100$\,keV, and likewise for 
Ge.%
\footnote{Atmospheric neutrinos are expected to begin contributing background events after about 50\,ton$\cdot$years of exposure \cite{neutrino}, and thus will not be an issue for this type of analysis for the foreseeable future.}
On the other hand we will take Ar experiments to have a low energy threshold of $30$\,keV.  Other parameters for these experiments, such as the energy resolutions, will also be taken from the same reference.%
\footnote{See also discussions in Refs\,\cite{Chang:2010yk,Feng:2011vu} 
for the prospects to test isospin-violating dark matter at future experiments.}

An issue in trying to constrain $f_p/f_n$ in practice is the uncertainty in the dark matter halo velocity distribution.  To the extent that different experiments probe different parts of halo phase space, a small change in the shape of the velocity distribution could be used to explain a given ratio of signal strengths, rather than a change in $f_p/f_n$.  As explained in \cite{Fox:2010bz}, the relevant issue is the range of recoil energies $[E_i, E_f]$ probed at each experiment, translated into a range of minimum velocities $[v_{\rm min,i }, v_{\rm min,f}] = [\sqrt{\frac{M_T E_i}{2 \mu_N^2}},\sqrt{\frac{M_T E_f}{2 \mu_N^2}}]$, with $M_T$ being the target mass.  Experiments with insufficient overlap in $v_{\rm min}$ space essentially cannot be used to robustly determine $f_p/f_n$ since they do not probe the same part of the velocity distribution.  For Xe, Ge and Ar, the $v_{\rm min}$ ranges probed are 
$60$--$190$\,km/s, $80$--$255$\,km/s and $190$--$345$\,km/s, respectively.  We thus see that Ar and Xe experiments will have essentially no $v_{\rm min}$ overlap, while Ar and Ge will only overlap in the higher energy Ge bins, where the Ge nuclear form factor will weaken any signal.  Moreover, the ratio of proton to neutron numbers in Ar (.82) and Ge (.78) only differ by about 5\%.  For this reason, Ar will not be a very useful element for an $f_p/f_n$ determination, unless the low energy threshold is able to be decreased.  On the other hand, Xe and Ge experiments will probe similar parts of the halo, and the proton to neutron ratio at Xe (.70) is about 10\% different from that of Ge.  In what follows, we will therefore only consider a comparison of Xe and Ge events.
%  Note that we will attempt to account for any remaining influence of halo uncertainty by considering various possible halo parameters during our analysis.

%One caveat, here, is that the range of the dark matter velocity for a given nucleus recoil energy 
%depends on the mass of the target elements.
%For example, the nucleus recoil energy in $10$\,keV to $100$\,keV corresponds 
%to the range of the minimal dark matter velocity $60-190$km/s for $^{131}$Xe and
%$78-250$km/s for $^{73}$Ge, respectively , while it is given by $180-340$km/s for $^{40}$Ar
%for the recoil energy in $30$\,keV to $100$\,keV.
%Thus, the velocity ranges in the $^{131}$Xe and $^{40}$Ar experiments barely overlap, 
%and hence, the signal ratio between the  $^{131}$Xe and $^{40}$Ar experiments 
%can be easily altered by a small change of the velocity distribution of dark matter,
%while the comparison between the  $^{131}$X and $^{73}$Ge experiments is expected to be 
%less sensitive to the velocity distribution.%
%\footnote{
%See Ref.\,\cite{Pato:2011de} for impacts of astrophysical uncertainties on the 
%measurement of dark matter properties in the case of light dark matter.
%}
%With this caution in mind, we discuss the testability of  the isospin violation 
%at  the next generation experiments by comparing the experiments with $^{131}$Xe
%and $^{73}$Ge.

In our analysis, we generate ``true" direct detection signals according to
 the cross section in Eq.\,(\ref{eq:DD}) for a given $M_{\rm DM}$ and for each target nucleus-- i.e. hypothetically what we are supposing each experiment will measure. 
For this purpose we have fixed the astrophysical parameters such as
the local dark matter density, $\rho_0 = 0.4$\,GeV/cm$^{-3}$,
the local circular velocity, 
$v_0 = 230$\,km/s,
and the Galactic escape velocity $v_{\rm esc} = 544$\,km/s assuming a Maxwell velocity distribution.
We then take the dark matter mass, dark matter neutron scattering cross section, circular velocity, and $f_p/f_n$ as free parameters, and compare with the signal recoil spectra of the ``true" signals  in 
ten linearly-spaced bins between $10$\,keV and $100$\,keV. 

\begin{figure}[tb]
\begin{center}
  \includegraphics[width=0.75\linewidth]{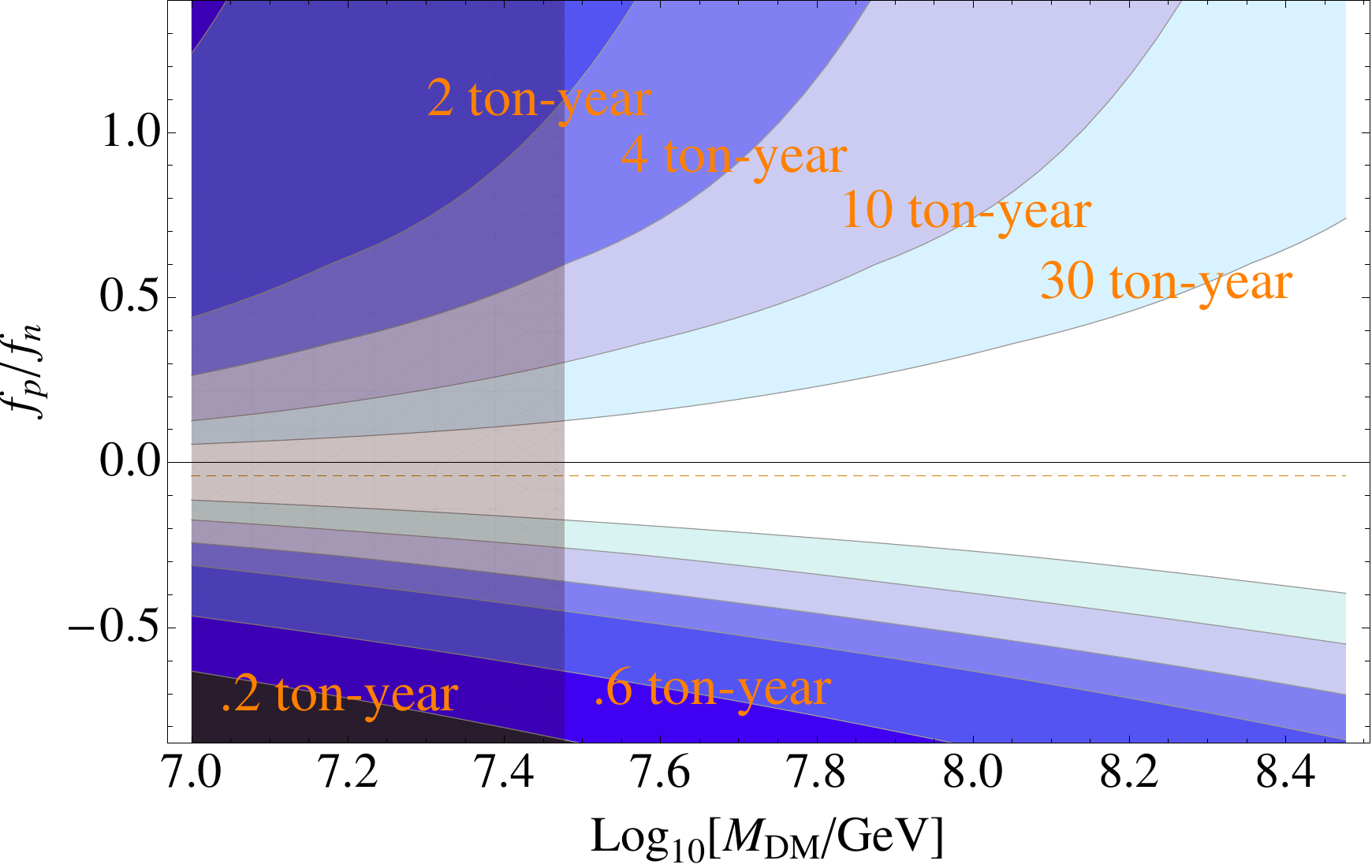}
\end{center}
\caption{\sl \small
The required effective exposures for a 90\%\,C.L. exclusion
of the isospin violation dark matter for a given hypercharged dark matter mass.
Isospin preserving dark matter corresponds to $f_p/f_n = 1$, and $Z$-mediated scattering is shown as the dashed line.
The shaded region has been excluded by the XENON100 experiments (see Eq.\,(\ref{eq:bound})).
}
\label{fig:exposure}
\end{figure}

In Fig.\,\ref{fig:exposure}, we show the required effective exposures at future Xe and Ge experiments (after cuts) for 90\%\,C.L. exclusions 
on values of $f_p/f_n$ for  given dark matter masses.
Here, we have marginalized
%the dark matter mass and
over the value of the dark matter neutron cross section, and the halo circular velocity.
Note that presently planned Xe and Ge experiments are expected to be able to achieve zero-background exposures of about 2\,ton$\cdot$years \cite{Pato:2011de}.
Interestingly, the figure shows that the isospin preserving hypothesis, i.e. $f_p/f_n = 1$, can 
be almost excluded at $90\%$\,C.L.  by multi ton-scale direct detection experiments (corresponding to a few hundred events)
for $M_{\rm DM} \lesssim 10^{8}$\,GeV.  This could arguably lend a fair bit of support to the hypercharged dark matter hypothesis, since it would give the simplest model to fit the data, with farther future experiments in principle able to strengthen the case.  Furthermore,  within the hypercharged dark matter framework, reproducing the observed signal strengths would fix the dark matter mass (to within a factor of $Y^2$) and probe the reheating epoch as we have discussed.\footnote{We also find that it would be possible to experimentally place lower bounds on the dark matter mass.  For example, if our scenario is realized in nature with $M_{\rm DM} \sim 10^{8}$\,GeV, then it should be possible to conclude that the dark matter is heavier than about $300$\,GeV through a similar analysis of direct detection data.}
%It should be noted that the above results are only mildly sensitive to the assumed halo velocity distributions, due to the large
%$v_{\rm min}$ overlap regions for Xe and Ge experiments.  We will leave further details of the impact of astrophysical uncertainties to future work.

%not so altered even if we change the astrophysical 
%parameters so much as long as we are assuming the Maxwell velocity distribution.
%The result is also expected to be less sensitive even to the small change of the shape of the velocity distribution 
%as long as we are comparing the signals in the $^{131}$Xe and $^{73}$Ge experiments% 
%\,\footnote{We will leave further details of the impact of astrophysical uncertainties to future work.}.

%%%%%%%%%%%%%%%%%%%%%%%%

%%%%%%%%%%%%%%%%%
\vspace{4pt}
{\bfseries\slshape  Discussion.}\/
To date,  no signs of new physics have appeared at the LHC, and it is appearing  likely that the hierarchy problem is at least incompletely solved, if not totally unsolved in nature.  
This diminishes the expectation for a TeV scale thermal relic associated with naturalness.
Even if the dark matter mass is heavy, and the relic abundance obtains an exponential sensitivity to $M_{\rm DM}/T_{\rm max}$, the associated fine tuning is actually rather mild-- of order 5\%.  It is even less severe if the reheating temperature is smaller than the dark matter mass, so that the Boltzmann factor is not the only effect suppressing the relic abundance.
Moreover, the coincidence between the sizes of $M_{\rm DM}$ and $T_{\rm max}$ may have a similar origin to the coincidence between the sizes of the cosmological constant and the energy density of the universe at the time of galaxy formation:  it has been argued that the dark matter density could be set by selection effects 
for the formation of structure and habitable planets, just as is the case for the CC \cite{anthropic}.  
If the dark matter mass is not forbidden by any symmetry, as in the minimal hypercharged scenario we have considered, then about a $5\%$ fine tuning may be completely irrelevant compared to the preference for the dark matter mass to be closer to the fundamental scale.

As a specific example, if supersymmetry 
is present in the fundamental theory, then its non-discovery at the LHC 
so far is suggestive that the supersymmetry breaking scale may be preferentially very high.  
In that case anthropic selection might set the supersymmetry breaking scale and/or the $\mu$ parameter to be appropriately close to $T_{\rm max}$ in order to yield an appropriate relic abundance for the 
lightest supersymmetric particle
(LSP).  Indeed, there are then two primary cases of interest: first,  it is possible that only the $\mu$ parameter is fine tuned to be close to $T_{\rm max}$ so that the Higgsino is perhaps alone at that scale.  A second possibility is that all of the superpartner masses are close to $T_{\rm max}$.  In this case the splitting between the neutral Higgsino components is of order $10\,{\rm keV}\times \frac{10^8 {\rm GeV}}{M_{\rm DM}}$.  Thus, if the Higgsino is the LSP, then in most of the mass range we have been considering the scattering will be effectively elastic, so that the signal will be as we have discussed.   It is then remarkable that these scenarios with very heavy superpartners become concretely testable and will be probed through several more orders of magnitude of interesting parameter space at upcoming experiments.

\section*{Acknowledgements} 
We would like to thank Shigeki Matsumoto and Natsumi Nagata for useful discussions.
This work is supported by the Grant-in-Aid for Scientific research from the
Ministry of Education, Science, Sports, and Culture (MEXT), Japan, No.\ 22244021 (TTY) and
No.\ 24740151 (MI).  It is also supported by the World Premier International Research Center Initiative (WPI Initiative), MEXT, Japan, and by STFC UK (BF).
%In that case, it is still possible that the dark matter particle may be selected from amongst the supersymmetric standard model candidates, with a mass which is simply fine tuned to be small. 
%It seems unlikely that a scalar superpartner would be chosen for this purpose, since the fine tuning required in the mass term is twice as severe as for the case of a fermion.  
%The viable possibilities are then the Higgsino, the wino and the bino.%
%\footnote{While the bino would not be in thermal equilibrium in the early universe if all of the other superpartners are very heavy, it may still be produced through the process Higgs + Higgs $\rightarrow$ bino + bino, through a virtual Higgsino.  This results in the relic abundance ??, which requires a similar degree of fine tuning as the other cases.}    
%While the wino or bino dark matter at $T_{\rm max}$ scale would likely be impossible to detect, 
%it is remarkable that the case of high scale Higgsino dark matter, requiring simply a fine tuning of the $\mu$ parameter, is actually highly testable, 
%and will be probed through several more orders of magnitude of interesting parameter space at upcoming experiments.

\end{document}